# Microscopic study of effective 2+1 dimensional gravity in ferrofluid-based hyperbolic metamaterials

Vera N. Smolyaninova [a], Jonathon Cartelli [a], Nathaniel Christopher [a], Benjamin Kist [a],

Jonathan Perry [a], Stephanie Spickard [b], Mary Sajini Devadas [b], Igor I. Smolyaninov [c]

[a] *Department of Physics Astronomy and Geosciences, Towson University,*

*8000 York Rd., Towson, MD 21252, USA*

[b] *Department of Chemistry, Towson University,*

*8000 York Rd., Towson, MD 21252, USA*

[c] *Department of Electrical and Computer Engineering, University of Maryland, College*

*Park, MD 20742, USA ; smoly@umd.edu*

Recent theoretical and experimental work demonstrated that nonlinear optics of ferrofluid-based hyperbolic metamaterials exhibits very unusual 2+2-dimensional spatiotemporal dynamics. Here we report a detailed microscopic study of mutual interactions of individual self-focused optical filaments inside this metamaterial. In agreement with theoretical expectations, the observed mutual interactions of individual filaments exhibit strong similarities with general relativity in 2+1 dimensions. This observation is very important since 2+1-dimensional gravity is an exactly solvable theory even in the quantum gravity limit.

Optical space is the central concept of the transformation optics paradigm [1]. A metric of optical space inside an electromagnetic metamaterial differs from the metric of physical space, and in such unusual cases as hyperbolic metamaterials [2], the optical space may behave like an "optical spacetime" [3]. Indeed, if we consider an extraordinary light signal propagating inside a hyperbolic metamaterial, the wave equation becomes

$$-\frac{\partial^2 \phi}{c^2 \partial t^2} + \frac{\partial^2 \phi}{\varepsilon_1 \partial z^2} + \frac{1}{\varepsilon_2}\left(\frac{\partial^2 \phi}{\partial x^2} + \frac{\partial^2 \phi}{\partial y^2}\right) = 0 \quad , \qquad (1)$$

where $\varepsilon_{xx} = \varepsilon_{yy} = \varepsilon_1$ and $\varepsilon_{zz} = \varepsilon_2$ are the dielectric tensor components of the metamaterial (calculated at the centre frequency $\omega$ of the signal bandwidth). The "wave function" of the extraordinary photon in Eq.(1) has been introduced as $\varphi = E_z$, where $E_z$ is the electric field component parallel to the optical axis $z$ of the metamaterial (see the detailed derivation in [3]). Depending on the composition and geometry, hyperbolic metamaterials may exhibit one or several frequency bands in which $\varepsilon_1$ and $\varepsilon_2$ change signs, and these signs may become opposite to each other [2]. In the particular case considered in [3], in which $\varepsilon_1 < 0$ and $\varepsilon_2 > 0$, Eq.(1) coincides with the Klein-Gordon equation for a scalar field $\varphi$ in a 2+2D spacetime having metric signature (-,-,+,+). It was noted that the nonlinear optical dynamics of this system should have very unusual two times (2T) character [3,4], and it may have strong resemblance with gravitational physics. Theoretical studies of the 2T space-time models had been pioneered by Dirac [5] and Sakharov [6]. More recent results may be seen in [7-9]. In particular, Bergshoeff and Sezgin extensively studied supergravity theories in 2+2 dimensions [7]. They have noted that the mathematics used in describing self-dual Yang-Mills or supergravity theories in 2+2 dimensions resembles that used in the discussion of Kaluza-Klein compactifications and solitonic solutions to string theories. While the most general case of 2+2D spatiotemporal dynamics merits further theoretical and experimental study, in several experimental situations it may be simplified and reduced to effective general relativity (GR) in 2+1 dimensions, which also appears to be very interesting. One of such situations involves using a monochromatic CW laser light illuminating a hyperbolic metamaterial, so that the first term in Eq.(1) is kept constant. Another such case may involve situations in which high power laser light passing through a hyperbolic metamaterial develops filamentation. If the filament radii remain approximately constant, the second term in Eq.(1) will be fixed and the nonlinear dynamics of light filaments will again look like gravity in 2+1 dimensions. Initial experimental evidence of such a filamentation inside a ferrofluid-based hyperbolic metamaterial has been communicated in [4]. Here we report a detailed microscopic study of this effect.

Before proceeding to experimental results, let us outline the basic features of the 2+1D situations mentioned above. First, let us consider a situation in which laser light of frequency $\omega$ passes through a hyperbolic metamaterial, and at the frequency of interest $\varepsilon_1(\omega) > 0$ and $\varepsilon_2(\omega) < 0$. In such a case the wave equation written for $\varphi_\omega$ becomes

$$-\frac{\partial^2 \phi_\omega}{\varepsilon_1(\omega)\partial z^2} + \frac{1}{\left(-\varepsilon_2(\omega)\right)}\left(\frac{\partial^2 \phi_\omega}{\partial x^2} + \frac{\partial^2 \phi_\omega}{\partial y^2}\right) = \frac{\omega_0^2}{c^2}\phi_\omega \quad , \tag{2}$$

and it may be interpreted as a wave equation for massive photons which live inside a 2+1D Minkowski spacetime in which the spatial coordinate $z$ behaves like a timelike variable. The nonzero metric coefficients $g_{ik}$ of this spacetime may be defined as

$$g_{00} = -\varepsilon_1(\omega) \qquad \text{and} \qquad g_{11} = g_{22} = -\varepsilon_2(\omega) \tag{3}$$

As demonstrated theoretically in [10] and experimentally in [4], the nonlinear optical Kerr effect leads to GR-like physics in this system. The Kerr effect must perturb this 2+1D Minkowski spacetime, resulting in self-focusing of the extraordinary light and formation of light filaments. This behaviour is consistent with the gravitational evolution of self-interacting scalar field in 2+1D found in [11]. When the filaments are formed, they must start to behave like world lines of massive point particles in 2+1D GR. In this limit the 2+1D spacetime in the source-free regions must become conformally flat [12], and the filament acceleration must reduce to

$$\ddot{x}^{\mu} + \Gamma^{\mu}_{\alpha\beta} \dot{x}^{\alpha} \dot{x}^{\beta} = 0 \quad , \tag{4}$$

where the derivatives are taken with respect to the timelike $z$ coordinate and $\Gamma^{\mu}_{\alpha\beta}$ are the Christoffel symbols. There is no long-distance interaction between static point particles. Indeed, since Einstein and curvature tensors are equivalent in three spacetime dimensions, localized sources only affect the global geometry, which is fixed by singularities of the worldlines of the particles [12].

On the other hand, if we are interested in the slow dynamics of the optical filaments as a function of physical time $t$, and the filament radii $\rho$ are approximately fixed, leading to fixed $k_z$:

$$\frac{k_z^2}{\varepsilon_1} \approx \frac{\omega^2}{c^2} - \frac{1}{\varepsilon_2}\left(k_x^2 + k_y^2\right) \approx \frac{\omega^2}{c^2} - \frac{1}{\rho^2 \varepsilon_2} \quad , \tag{5}$$

we can use the Born–Oppenheimer approximation to separate the fast and slow variables in Eq.(1), so that the wave equation for the slow variables becomes

$$-\frac{\partial^2 \phi}{c^2 \partial t^2} + \frac{1}{\varepsilon_2}\left(\frac{\partial^2 \phi}{\partial x^2} + \frac{\partial^2 \phi}{\partial y^2}\right) = \frac{k_z^2}{\varepsilon_1} \phi \quad , \tag{6}$$

where both $\varepsilon_1$ and $\varepsilon_2$ must be positive at near zero frequencies of interest for the slow dynamics. The spatiotemporal dynamics of the optical field becomes 2+1D, and the filament interaction once again is defined by Eq.(4), where the derivatives are now taken with respect to the physical time $t$.

Let us now describe how this theoretical picture holds in the experiment. While the initial observations of effective gravity in ferrofluid-based hyperbolic metamaterials demonstrated such interesting 2T dynamical effects as self-focusing and filamentation of the initially Gaussian laser field [4], the spatial resolution of these experiments was not sufficient to reliably study individual optical filaments and their mutual interactions.

In this work we substantially improved the spatial resolution of our experimental setup, so that the properties of individual filaments and details of their mutual interaction have been revealed in fine detail. The schematic diagram of our experimental setup is shown in Fig. 1. Compared to the original setup described in detail in [4], we have implemented a high-resolution telescopic objective mounted on the LWIR camera. We have studied propagation of an initially Gaussian horizontally polarized $CO_2$ laser beam operating at the 10.6 $\mu$m wavelength through a ferrofluid-based hyperbolic metamaterial sample contained in a thin (100 $\mu$m) optical cuvette. A LWIR camera with an attached telescopic objective was used to image the beam shape after its passage through the metamaterial. The self-assembled ferrofluid-based hyperbolic metamaterial [13] inside the cuvette was formed under the influence of external DC magnetic field. Since magnetism of the iron-cobalt nanoparticles in the ferrofluid is rather weak, in the absence of external magnetic field the nanoparticles are randomly distributed within the fluid (kerosene). Application of a modest (~ 300 Gauss) external magnetic field leads to formation of metal nanocolumns inside the fluid, so that a self-assembled hyperbolic metamaterial is formed, which optical axis is directed along the field [13]. Depending on the orientation of the magnetic field with respect to the cuvette, the laser light passing through the metamaterial becomes either ordinary (E field of photons oriented perpendicular to the optical axis), or extraordinary (E field of photons has a non-zero component parallel to the optical axis). In agreement with theoretical expectations [10, 14-16], strong nonlinear effects in laser beam propagation were observed only for the extraordinary state of light polarization [4]. While the ordinary light beam shape hardly experienced any changes over time, the extraordinary beam separated into multiple filaments, which exhibited fast dynamical behaviour on the millisecond time scale, as shown in videos in Ref. [4]).

An example of high-resolution image of the extraordinary laser beam propagated through the metamaterial sample is shown in Fig. 2. This 6.48 mm x 6.48 mm image frame was taken from a recorded video of the extraordinary beam evolution as a function of time. This image was recorded at 960 mW $CO_2$ laser power. The average (approximately Gaussian) laser beam profile was subtracted from the image. This background subtraction allowed to reveal the optical filaments formed within the beam and study the temporal dynamics of their mutual interactions. The filaments appeared to be quite stable, having an average lifetime of about 10 s. At higher laser powers, when the filaments appeared to be well-defined, their width to intensity ratio was approximately constant (Fig.2b), which justifies the description of filament dynamics using Eq.(6). The increase of laser power passing through the metamaterial led to increase of the number of filaments (Fig.2c), while their relative sizes remained mostly similar. This initial increase was followed by eventual near saturation of the number of filaments at higher powers.

An example of temporal evolution of the positions of multiple filaments within a 2 x 2 mm$^2$ area of the laser beam may be seen in Fig.3a. Each filament is represented by a different colour. The average light intensity within a filament is represented approximately via its point sizes. As can be seen from this plot, while some filaments do not exhibit any clear sign of mutual interaction, several filaments in the field of view interact with each other. In particular, several filaments fall onto the most intense filament within the field of view (shown by cyan colour), and they are absorbed by this filament. As demonstrated by Figs.3b-d, this event may be used to evaluate the distance dependence of the inter-filament forces. Fig.3b shows the measured distances between the three filaments and the filament shown by cyan colour as a function of time. The same colour coding is used in Figs 3a and 3b. The smoothed data for the distance between the pink and cyan filaments shown in Fig.3c (blue line) was used to calculate acceleration of the pink filament as a function of distance between these filaments (Fig.3d). Based on these measurements, we may conclude that there is no substantial long-distance force acting between the filaments. The force appears to be contact in nature. It manifests itself only when the inter-filament distance becomes comparable with the filament size. This character of the inter-filament interaction is consistent with expectations based on the analogous 2+1D gravity. As mentioned above, the 2+1D optical spacetime in the source-free regions must be conformally flat [12], and the expected inter-filament interaction must be mainly topological in nature. Indeed, the velocity-dependent "gravimagnetic-like" forces described by Eq. (4) appear to be too small to be detectable with the currently achieved precision of our experiments, since they are proportional to a factor of the order of $v^2/c^2$.

Let us now consider the expected topological interactions of filaments within the scope of the effective 2+1D gravity model [12]. We will limit our consideration to the zero angular momentum $m=0$ ground state case, for which the extraordinary (TM) and ordinary (TE) polarization states of light inside the filaments remain well-defined. The peculiar nature of filament interactions is illustrated in Fig.4. While the optical spacetime remains locally flat, an effective point mass $M_l \sim I_l$ (where $I_l$ is the light intensity inside the filament) excludes an angular sector (shown in grey) from the flat 2D optical space around it, as illustrated in Fig.4a. As a result, the optical spacetime near the filament becomes conical [12] and its metric is given by

$$ds_{opt}^2 = -d\tilde{z}^2 + d\tilde{r}^2 + \alpha^2 \tilde{r}^2 d\phi^2 = -d\tilde{z}^2 + d\tilde{r}^2 + \tilde{r}^2 d\tilde{\phi}^2 \ , \tag{7}$$

where $d\tilde{z}^2 = \varepsilon_1 dz^2$ is the time-like variable, $d\tilde{r}^2 = -\varepsilon_2 dr^2$ (see Eq.(2)), and the range of $\tilde{\phi}$ is smaller than $2\pi$. Note that the metamaterial examples of conical optical spaces may be also found in [17,18]. The components of the anisotropic dielectric tensor around the filament may be found from Kerr effect as

$$\varepsilon_{ij} = \varepsilon_{ij}^{(0)} + \chi_{ijlm}^{(3)} E_l E_m \tag{8}$$

For the extraordinary ($E_z \neq 0$) light filaments the sign of effective gravity is defined by the sign of $\chi^{(3)}_{zzzz}$. In agreement with [10], negative sign of $\chi^{(3)}_{zzzz}$ leads to $\left|\varepsilon_{zz}\right| > \left|\varepsilon^{(0)}_{zz}\right|$, which means that $\alpha = \dfrac{P}{2\pi\tilde{r}} < 1$ , where $P$ is the perimeter of a circle of radius $\tilde{r}$ in the optical space, which remains constant far from the filament. As a result, the sign of effective gravity is attractive.

Since the exact magnitude of $\chi^{(3)}_{zzzz}$ in the ferrofluid-based hyperbolic metamaterial is not known, we may base our numerical estimates on the magnitude of the thermo-optical nonlinear refractive index $n_2 = 7.4 \times 10^{-8}$ cm$^2$/W which was measured in a somewhat similar suspension of gold nanoparticles in castor oil [19]. Based on Eq.(7) and reference [12], the cone angle $\beta_1$ equals $\pi(1-\alpha) = 4\pi G M_1 \sim 4\pi n_2 I_1$, where $G$ is the effective gravitational constant, which is proportional to $\chi^{(3)}_{zzzz}$, and therefore, to the nonlinear refractive index $n_2$ of the medium [4,10] (the thermo-optical nature of the nonlinearity leads to $\chi^{(3)}_{zzrr} << \chi^{(3)}_{zzzz}$ in metal nanoparticle chains). Note that the red dots shown in the figure must be identified with each other. Addition of a point mass $M_2$ at some distance from $M_1$ leads to exclusion of an additional angular sector from the optical space, as indicated in Fig.4b, so that the full angle is $\beta_1 + \beta_2 = 4\pi G(M_1 + M_2)$. The red and the yellow dots (and the seemingly two separate locations of $M_2$ in the physical space) must be identified with each other in the optical space.

This topological picture of filament interactions has two important observational consequences. Since the picture of filaments is being observed in the physical space, the observation of sudden jumps in filament positions is to be expected, since the two filaments observed separately may in fact be one and the same filament in the optical space. Alternatively, the two filaments which appear distant in the physical space may in fact be very close to each other in the optical space, and therefore they may interact very strongly with each other.

Another observational consequence of this model is that because the total cone angle $\beta$ of a large number of filaments cannot exceed $\pi$ in the optical space, the total number of filaments cannot be arbitrary large without closing the optical space around the filaments [12], or the necessity of negative $M$ filaments (dark filaments) to appear in the system (note that in the language of transformation optics such a closed optical space is considered to be "cloaked" [1]). Quite obviously, the same topological limit on the total number of filaments must exist in both $xyz$ (see Eq.(2)) and $xyt$ (see Eq.(6)) 2+1D spacetimes. The fact that the number of filaments indeed saturates at higher optical power (see Fig.2c) may be considered as evidence of such a topological limit (note that the total optical power in the beam may be redistributed between the overall Gaussian beam and the filaments). Eqs.(5,6) from [4] indicate that because of the linear alignment of the ferrofluid nanoparticles in magnetic field, the nonlinear refractive index of the ferrofluid is amplified by the factor of $\varepsilon_2 \approx$ -100 (see Fig.3c from [13]). This should bring the magnitude of the nonlinear refractive index to about $n_2 \sim 10^{-5}$ cm$^2$/W. The critical power leading to self-focusing of the optical filaments in the ferrofluid-based hyperbolic metamaterial has been estimated in [4] as $P_{cr} \sim 3$ mW. Assuming the same $P_{cr}$ optical power being trapped in each filament, the topological limit on the number of filaments $N_{top}$ may be estimated as

$$4\pi n_2 \frac{P_{cr}}{r^2} N_{top} \approx \pi \ , \qquad (9)$$

where $r$ is the filament radius. This estimate leads to

$$N_{top} \approx \frac{r^2}{4n_2 P_{cr}} \approx \left(\frac{r}{3\mu m}\right)^2 \tag{10}$$

Assuming based on Fig.2c that $N_{top}\sim100$, the "standard filament" radius should be $r\sim30\mu m$, which is not far from our experimental observations. While these estimates are indeed very rough, they seem to be consistent with the effective 2+1D gravity interpretation of our experiments. We should also note that experimentation with more complex $m\neq0$ nonlinear filament systems may reveal more sophisticated features of the effective optical spacetime of this model, such as the jump property of the coordinate time which was noted in [12]. In addition, it should be interesting to study this or similar systems in the quantum mechanical regime, so that a direct comparison can be made with 2+1D quantum gravity, which is known to be an exactly solvable theory [20].

## AKNOWLEGEMENTS

The authors would like to acknowledge W. Korzi and J. Klupt for experimental and technical help. This work was supported in part by FDRC, SET and Fisher Endowment grants at Towson University.

**Figure Captions**

**Figure 1.** Schematic diagram of the experimental setup. A LWIR camera (FLIR Systems) with an attached telescopic objective is used to study propagation of $CO_2$ laser beam through the ferrofluid-based hyperbolic metamaterial, which is subjected to external DC magnetic field. The inset shows the measured beam shape of the $CO_2$ laser in the absence of the metamaterial sample. Two orientations of the external magnetic field $B$ used in our experiments are shown by green arrows. The red arrow shows polarization of laser light.

**Figure 2.** (a) A 6.48 mm x 6.48 mm image frame taken from a recorded video of the extraordinary beam evolution as a function of time. This image was recorded at 960 mW laser power. The average (approximately Gaussian) laser beam profile was subtracted from the image to reveal optical filaments within the beam. (b) At higher laser powers, when the filaments appeared to be well-defined, their width to intensity ratio is approximately constant. (c) The increase of laser power passing through the metamaterial leads to increase of the number of filaments. The line is a guide to the eye.

**Figure 3.** (a) Temporal evolution of multiple filaments within a ~ 2 x 2 mm$^2$ area of the laser beam. Each filament is represented by a different color. The average light intensity within a filament is represented via the size of the points. (b) Measured distance between filaments as a function of time for the four filaments which appear to merge. The same color table is used. (c) Distance between the pink and the cyan filaments as a function of time. Smoothed interpolation of the data is shown by blue line. (d) Calculated acceleration as a function of distance between the filaments shown in (c).

**Figure 4.** (a) A point effective mass $M_1$ excludes an angular sector (shown in grey) from the flat two-dimensional optical space around it, so that the optical space becomes

conical. The cone angle $\beta_1$ equals $4\pi GM_1$. The red dots must be identified with each other. (b)  Addition of point mass $M_2$ leads to exclusion of additional angular sector from the optical space, so that the full angle becomes $\beta_1 + \beta_2 = 4\pi G(M_1 + M_2)$.  The red and the yellow dots (and the seemingly two separate locations of $M_2$) must be identified with each other.

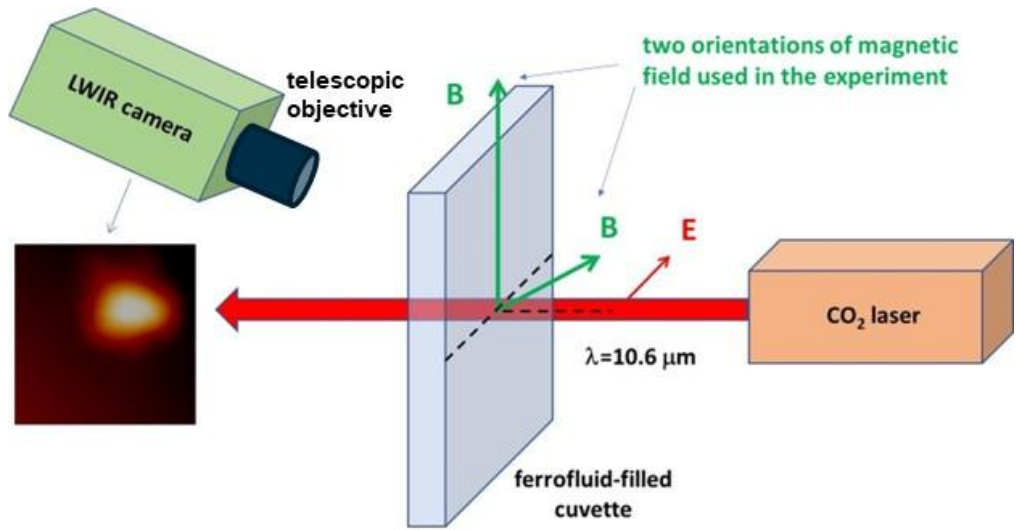

**Fig. 1**

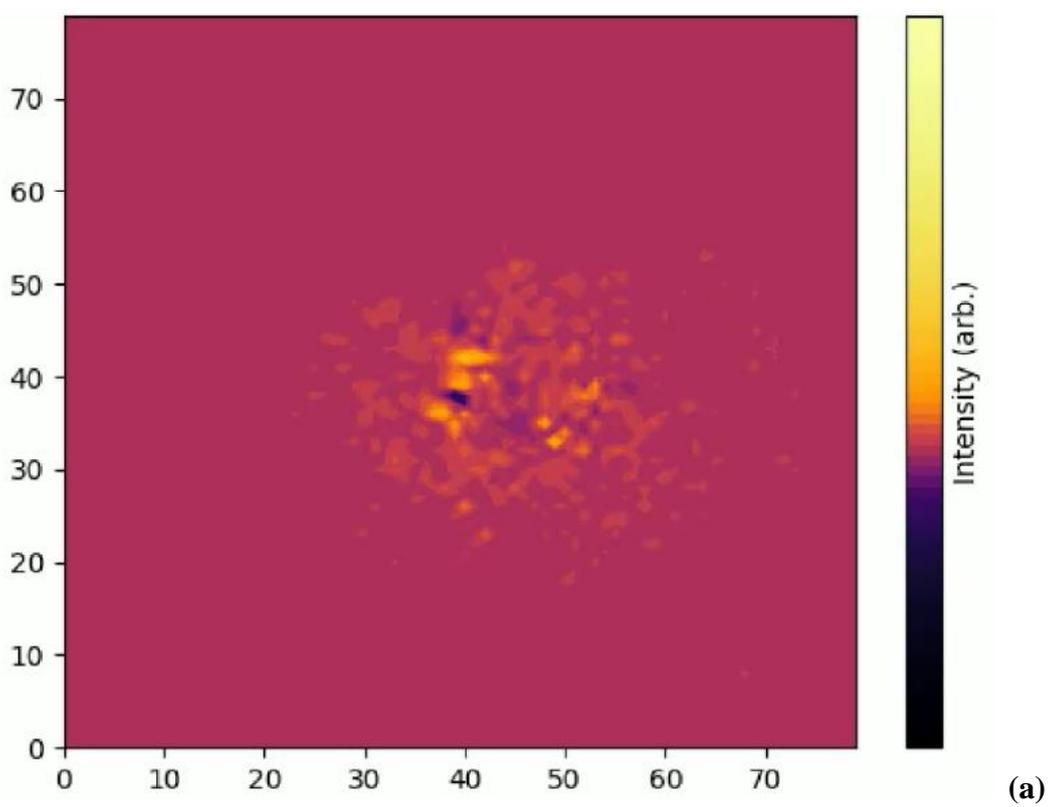

**(a)**

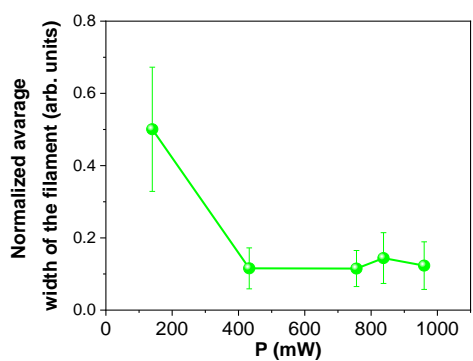

**(b)**

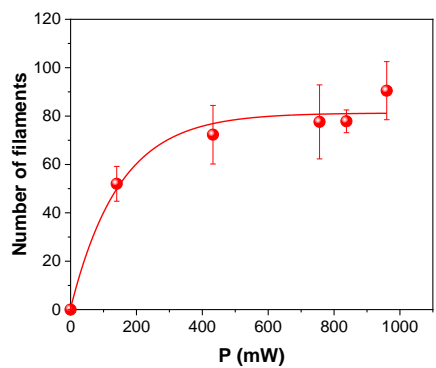

**(c)**

**Fig. 2**

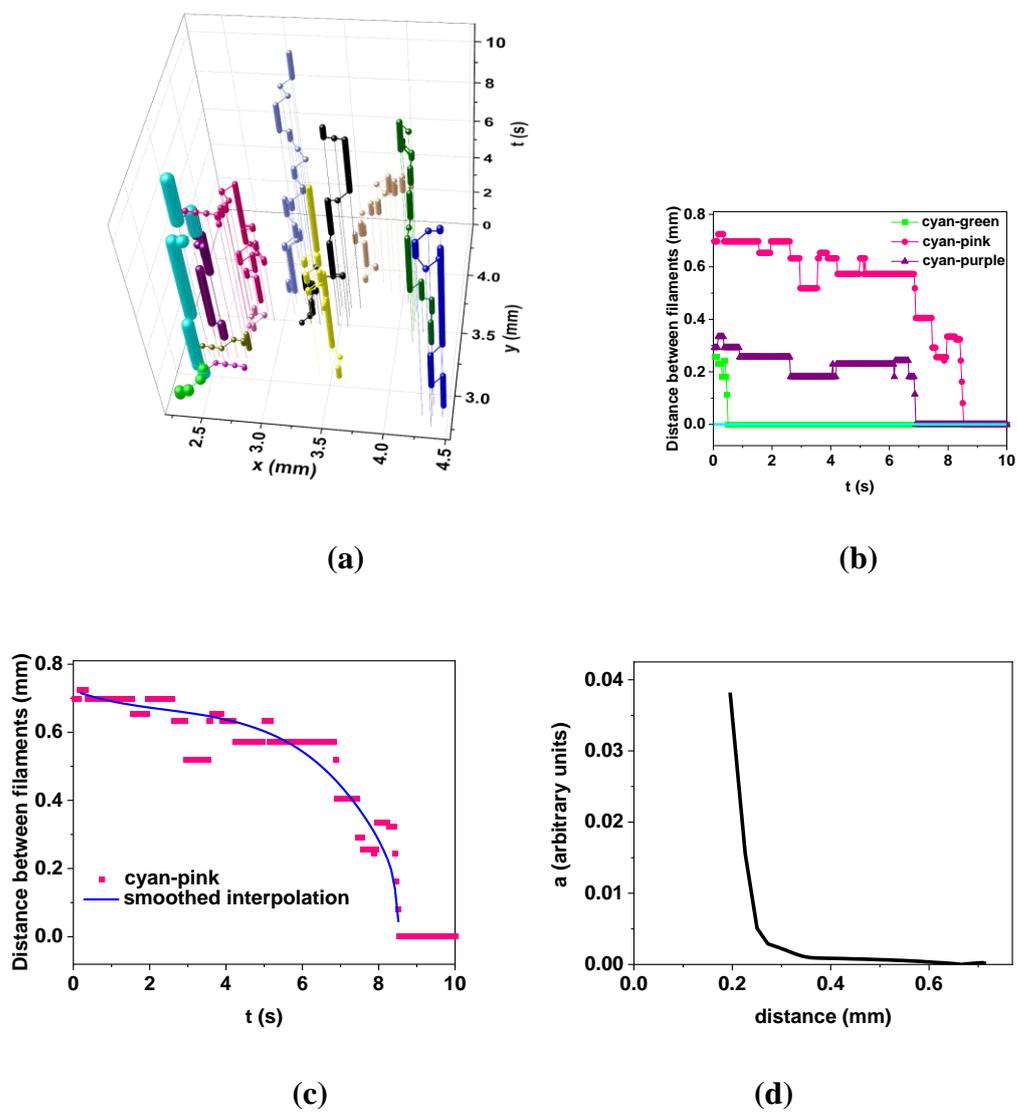

(a)

(b)

(c)

(d)

**Fig. 3**

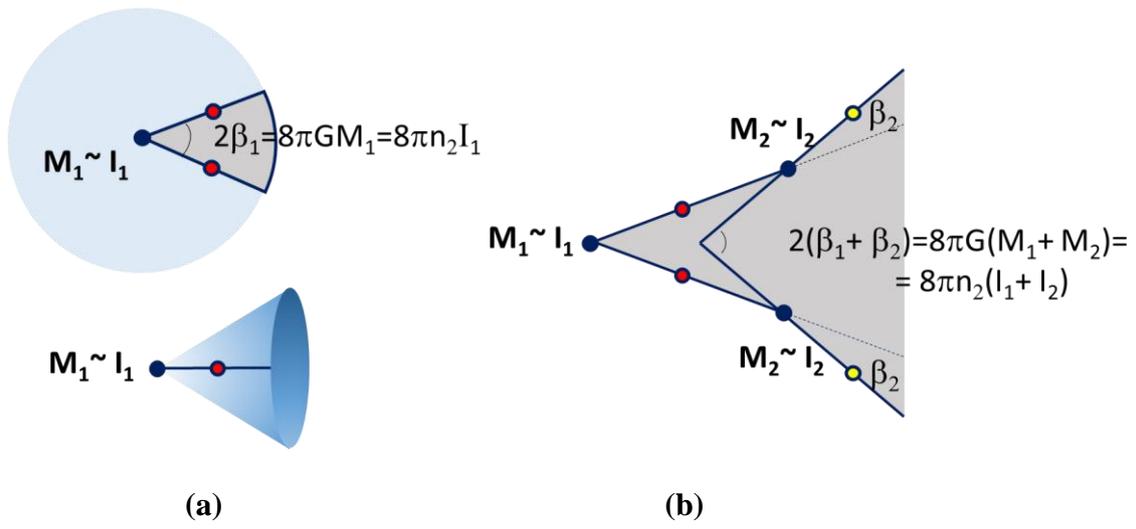

$M_1 \sim I_1$

$2\beta_1 = 8\pi GM_1 = 8\pi n_2 I_1$

$M_1 \sim I_1$

$M_2 \sim I_2$   $\beta_2$

$M_1 \sim I_1$

$2(\beta_1 + \beta_2) = 8\pi G(M_1 + M_2) = 8\pi n_2(I_1 + I_2)$

$M_2 \sim I_2$   $\beta_2$

**(a)**        **(b)**

**Fig. 4**